\documentclass[a4paper,11pt]{article}
\usepackage{pos}
\usepackage{wrapfig}

\newcommand\tcwidth{.16\linewidth}

\title{Ultrafast CMOS image sensors and data-enabled super-resolution for multimodal radiographic imaging and tomography}
\author[a,\dagger]{Xin Yue}
\author[b,c,\dagger]{Shanny Lin}
\author[b]{Wenting Li}
\author[b] {Bradley T. Wolfe}
\author[b]{Steven Clayton}
\author[b]{Mark Makela}
\author[b]{C. L. Morris}
\author[d]{Simon Spannagel}
\author[e]{Erik Ramberg}
\author[e]{Juan Estrada}
\author[c]{Hao Zhu}
\author[a]{Jifeng Liu}
\author[a]{Eric R. Fossum}
\author*[b]{Zhehui Wang}

\affiliation[a]{Thayer School of Engineering at Dartmouth, Dartmouth College,\\
  Hanover, NH 03755, USA}

\affiliation[b]{Los Alamos National Laboratory,\\
Los Alamos, NM 87545, USA}

\affiliation[c]{Chandra Family Department of Electrical \& Computer Engineering, The University of Texas at Austin,\\
Austin, TX, 78712, USA}

\affiliation[d]{Deutsches Elektronen-Synchrotron DESY, \\
Notkestr. 85, 22607 Hamburg, Germany}

\affiliation[e]{Fermi National Accelerator Laboratory,\\
Batavia, IL 60510, USA}

\affiliation[\dagger]{Lead authors with equivalent contributions}

\emailAdd{zwang@lanl.gov}
\abstract{We summarize recent progress in ultrafast Complementary Metal Oxide Semiconductor (CMOS) image sensor development and the application of neural networks for post-processing of CMOS and charge-coupled device (CCD) image data to achieve sub-pixel resolution (thus `super-resolution'). The combination of novel CMOS pixel designs and data-enabled image post-processing provides a promising path towards ultrafast high-resolution multi-modal radiographic imaging and tomography applications. }

\FullConference{%
  10th International Workshop on Semiconductor Pixel Detectors for Particles and Imaging (Pixel2022)\\
  12-16 December 2022\\
  Santa Fe, New Mexico, USA}

\begin{document}
\maketitle

\section{Introduction}

Using X-rays for imaging and tomography of optically opaque objects dated back to the famous invention of Wilheim R\"ontgen in 1895. Multimodal (MM) radiographic imaging and tomography (RadIT) combines several different forms or energies of ionizing radiation such as X-rays, $\gamma$-rays, neutrons, energetic electrons, protons and others, which can potentially yield more information about an object than by using a monochromatic (mono-energetic) photons (particles) alone as the source of illumination.

\begin{figure}[!htb]
\centering
\includegraphics[width = 0.95\linewidth]{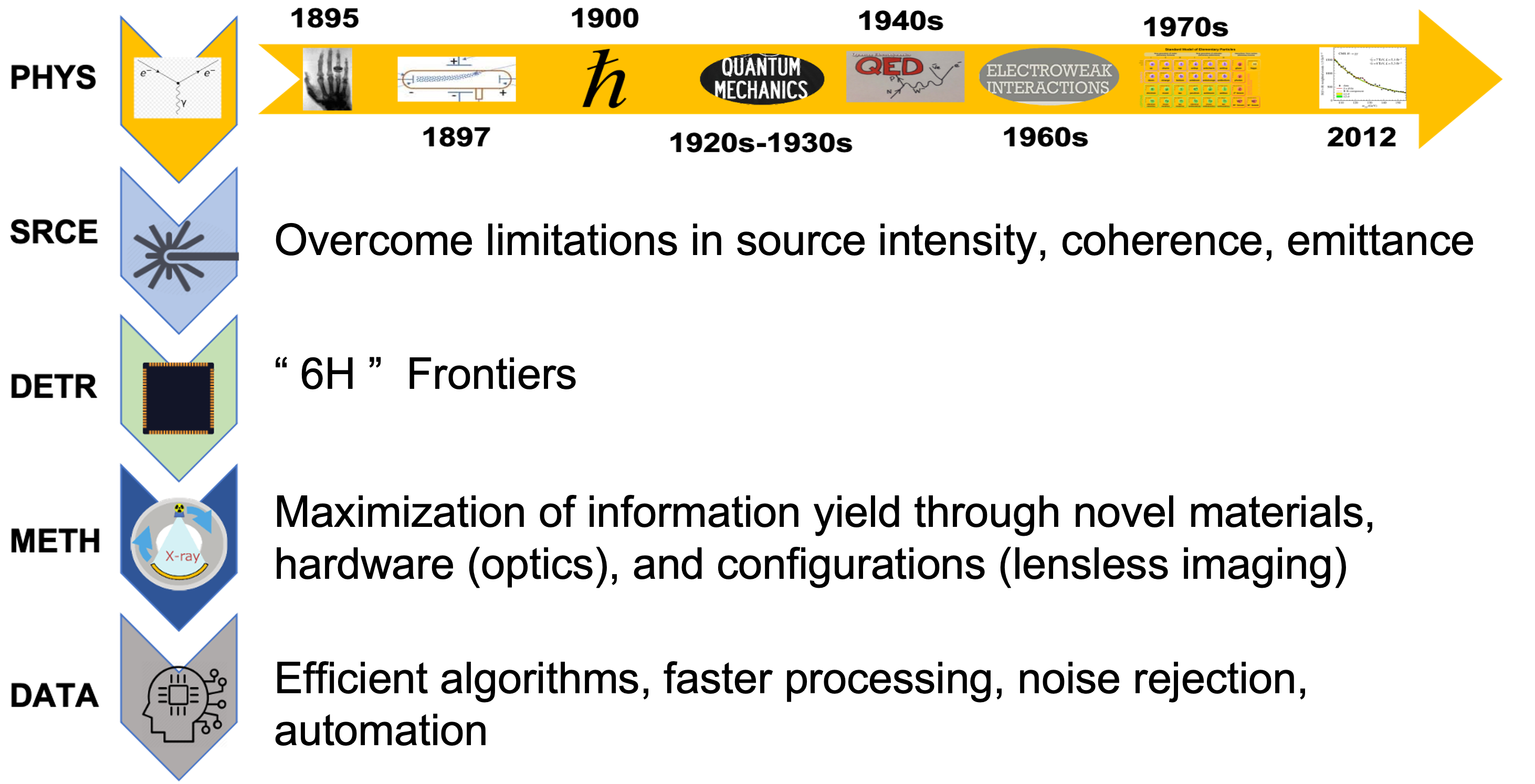}
\caption{A holistic approach to MM RadIT optimization includes at least five branches of effort: Fundamental physics of radiation-interaction with matter (PHY), radiation sources (SRCE), detectors (DETR), methods to modulate the radiation field (METH) and data handling (DATA). The fundamental physics principles of MM RadIT are well established.  Some challenges and opportunities for SRCE, DETR, METH and DATA are listed above for each branch.}
\label{fig:Hol1}
\end{figure}

The technical challenges and opportunities for MM RadIT can be summarized in Figure~\ref{fig:Hol1}. Additional information may be found in~\cite{Wang:2022}. The physics framework of radiation-matter interactions is mostly complete for all practical purposes. Recent advances in high-intensity radiation sources such as synchrotrons, X-ray free electron lasers, high-current low-emittance charged particle accelerators, laser-driven sources open door to MM RadIT. One of the optimization problems in MM RadIT is to obtain as high temporal, spatial resolution of an object as possible, with a sufficiently large field of view, and at a certain radiation dose to minimize radiation damage of the object. 

Below, we first describe the recent progress in ultrafast Complementary Metal Oxide Semiconductor (CMOS) pixelated sensor design and prototyping, followed by the the use of neural networks for noise emulation in X-ray imaging, and the demonstration of sub-pixel resolution in neutron detection. Follow-on work includes CMOS image sensor fabrication and extension of the neural networks to different types of particles or photons, and different noise environment. Our work demontrates a promising path towards high-speed high-resolution multi-modal radiographic imaging and tomography applications. 

\section{Ultrafast CMOS image sensor development}

\begin{figure}[!htb]
\centering
\includegraphics[width = 0.8\linewidth]{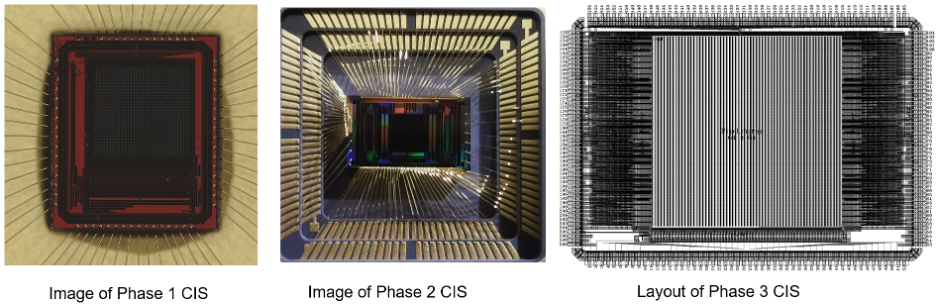}
\caption{Phase1, 2, 3 CMOS image sensor (CIS) taped out in this research.}
\label{fig:Dx}
\end{figure}

Ultra-high-speed (UHS) or ultrafast image sensors are widely used in scientific and industrial applications. The research on UHS CMOS image sensors for X-ray regimes has been conducted for years. The recently published works~\cite{Cro:2013, Suz:2020} push the frame rate of UHS image sensors to the range of millions of frames per second (Mfps) by adopting burst-mode operations and advanced CMOS technology or customized CMOS technology. Three CMOS image sensors were taped out towards this goal, as shown in Figure~\ref{fig:Dx}.

During the phase one of this research~\cite{Wan:2021}, theoretical modeling and preliminary tests demonstrated more than 10$\times$ quantum efficiency improvement for high-energy X-ray photons ($>$10 keV) by depositing a photon-attenuation-layer (PAL) on a CMOS image sensor~\cite{Eld:2021, Kat:2021}. In the phase two of this research, a block-wise compact readout architecture based on unit-length-capacitor and asynchronous successive-approximation (SAR) analog to digital converter (ADC)~\cite{Har:2019} was proposed and implemented, which enabled the image sensor fabricated using a standard 180-nm process to run at 76 thousand-frames-per-second (kfps). To further boost the frame rate of the image sensor, a burst mode image sensor based on sequential transfer gates was proposed and taped out in Oct. 2022 during the phase three of this research. The sequential transfer gates enable the image sensor to run at least 20 Mfps and achieve the lowest input-referred noise. Some highlights of this burst-mode image sensor is included below.

\begin{figure}[!htb]
\centering
\includegraphics[width = 0.4\linewidth]{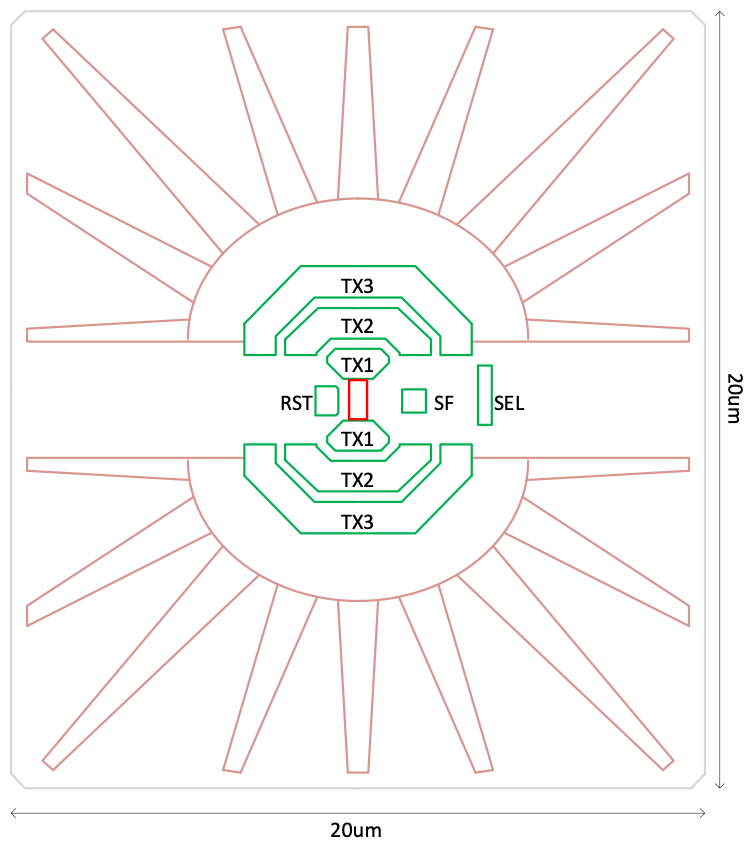}
\caption{High-Speed Conceptual Layout of a pixel in a CMOS camera. See the text for further details.}
\label{fig:dart1}
\end{figure}

Figure~\ref{fig:dart1} shows the conceptual 20-\textmu m pixel layout based on the sequential transfer gates, where the orange rising-run shape stands for the photodiode, and the green polygon stands for the transfer gate. Each photodiode finger geometry shape has been carefully calculated and optimized to have a constant $\sim$800V/cm strong electrical field pointing from the tip of the photodiode to the center of the photodiode, which guarantees fast charge transfer without process modification. 

\begin{figure}[!htb]
\centering
\includegraphics[width = 0.95\linewidth]{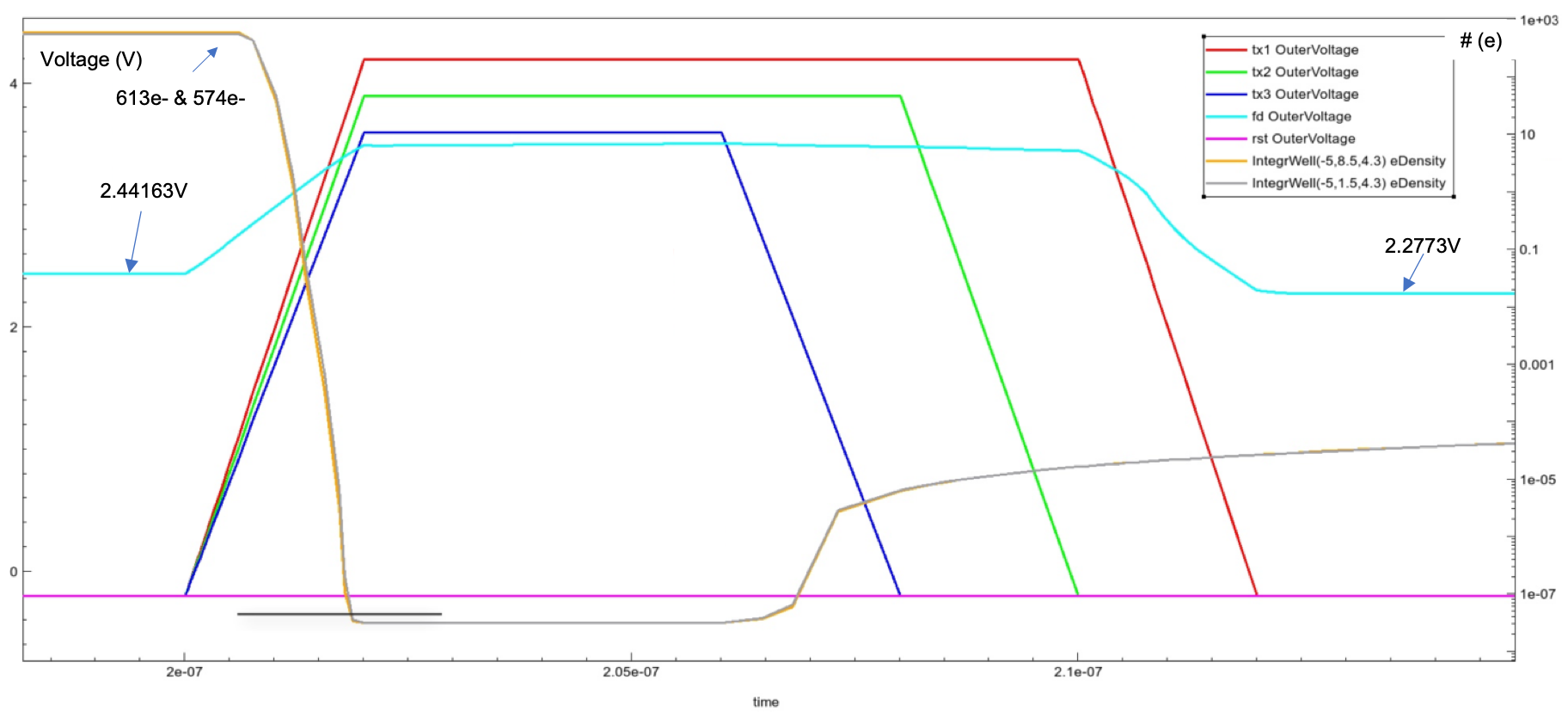}
\caption{An example of TCAD transient simulation during charge transfer. The text contains further details.}
\label{fig:dart2}
\end{figure}

As Figure~\ref{fig:dart2} shows, by applying monotonically increasing control voltages and sequential timing on TX3(Blue), TX2(Green), and TX1(Red)  gates, electrons (purple and cyan) in photodiodes can be fully transferred within 12 ns, with no image lag noticed in simulation. 

\begin{figure}[!htb]
\centering
\includegraphics[width = 0.95\linewidth]{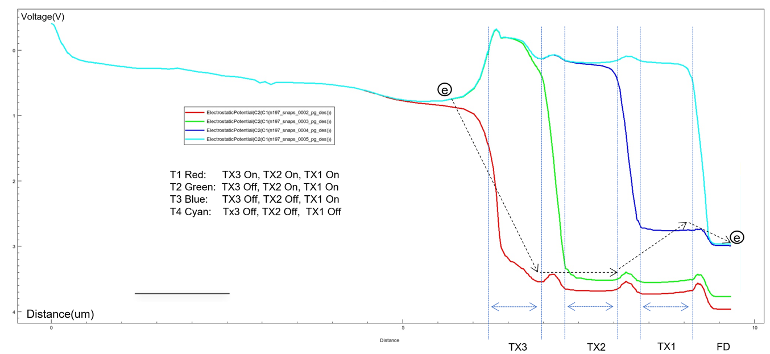}
\caption{Electrostatic potential along the charge transfer path.}
\label{fig:dart3}
\end{figure}

Figure~\ref{fig:dart3} shows the potential diagram during the charge transfer path. One can see that photon-generated electrons are first swept into the channel under the TX3 gate due to the strong electrical field in photodiode fingers. Then TX3, TX2, and TX1 gates turn off sequentially, which pushes electrons to move toward the floating diffusion node. Because the TX2 gate is entirely off before the falling transition of the TX1 gate, it is safe to move the floating diffusion node away from TX1, which will effectively reduce the overlap capacitance between the TX1 gate and floating diffusion node, increase the conversion gain of the pixel and reduce input-referred noise. At the same time, the electrical field between the TX1 gate and the floating diffusion node is reduced, which also reduces the dark current due to Gate-Induced-Drain-leakage (GIDL). A test chip is designed based on this sequential transfer gate pixel in a standard 180-nm process without any process modification. Simulations show that the test chip can run at least 20 Mfps with less than 5.8 $e^-$ input-referred noise, the lowest noise reported in the ultrafast burst-mode image sensor category. 

\section{Noise emulation using neural network}
\begin{figure}
  \centering
  \begin{tabular}{cccc}
    \includegraphics[width=\tcwidth]{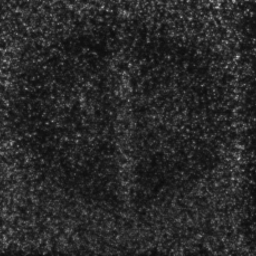} &   \includegraphics[width=\tcwidth]{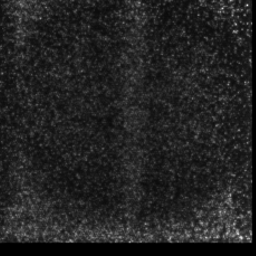} & \includegraphics[width=\tcwidth]{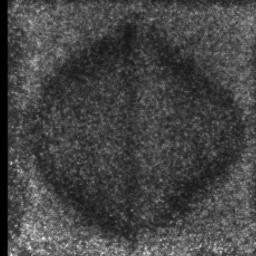} & \includegraphics[width=\tcwidth]{ 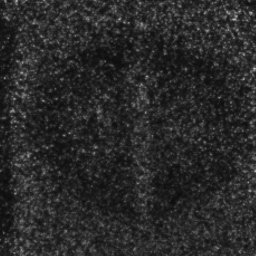} \\
    \includegraphics[width=\tcwidth]{ 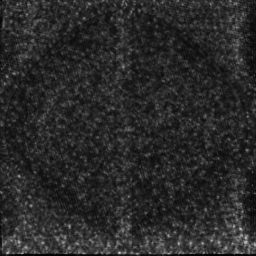} &   \includegraphics[width=\tcwidth]{ 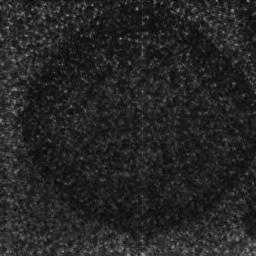} & \includegraphics[width=\tcwidth]{ 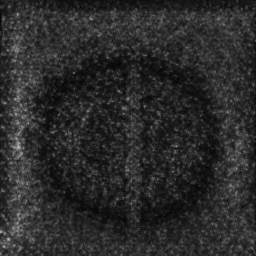} & \includegraphics[width=\tcwidth]{ 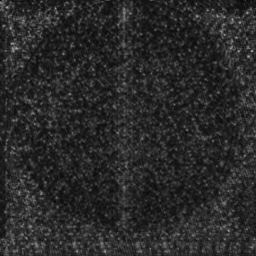} \\
  \end{tabular}
  \caption{Comparison of experimental images (Top Row) and  images produced using Contrastive Unpaired Translation (Bottom Row).}\label{fig:syn}
\end{figure}

Noise is ubiquitous in imaging and especially in high-speed and ultrafast imaging, when the signal-to-noise ratio is limited in part by the source intensity and transients that may be induced in the electronics. Meanwhile, better understanding of the noise through modeling is useful for noise reduction. Some examples of noisy images from inertial confinement fusion (ICF) experiments are shown  in Figure~\ref{fig:syn}. Further details may be found for example in~\cite{Wolfe:2021}. Here we describe the use of a generative adversarial network (GAN) to emulate image noise from the experiments. Some examples of the synthetic images are given in Figure~\ref{fig:syn}, which are qualitatively similar to the experimental data.

\begin{figure}[!htb]
\centering
\includegraphics[width = 0.75\linewidth]{ 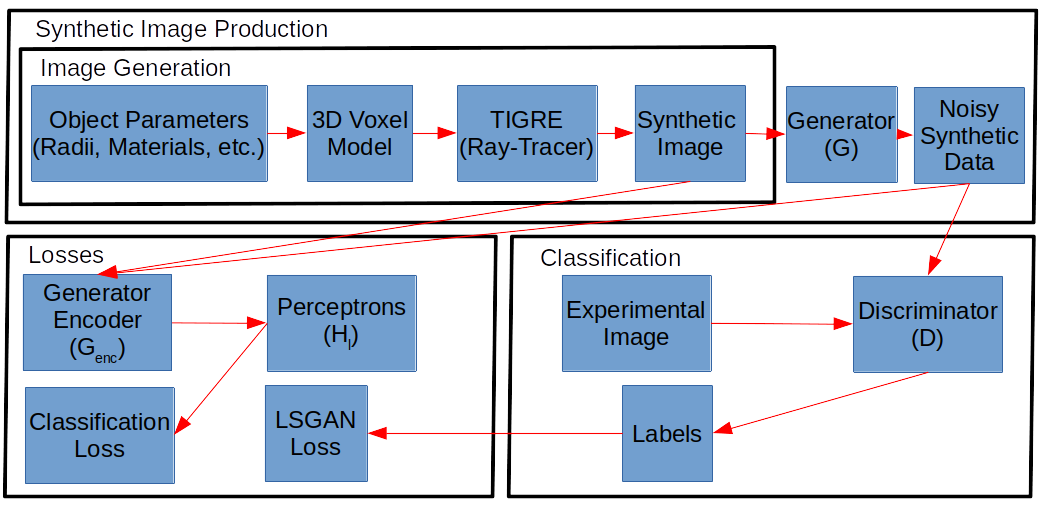}
\caption{The process of training CUT using synthetically produced radiographs and experimental images}
\label{wrap-fig:cut}
\end{figure}

%\begin{wrapfigure}{r}{5.5cm}
%  \caption{The process of training CUT using synthetically produced radiographs and experimental images.}\label{wrap-fig:cut}
%  \includegraphics[width=6.0cm]{ flowchart.png}
%\end{wrapfigure}
The work flow of image synthesis is summarized in Figure~\ref{wrap-fig:cut}. The first step is to produce noise-free synthetic radiographs. Three dimensional (3D) models of ICF shells are generated using Legendre polynomials for the shell boundaries and constant densities for the shells.
These shells consist of a $SiO_2$ inner shell, an aluminum outer shell, and a foam between the two shells~\cite{Wolfe:2021}.
These computer generated 3D shell models are projected to 2D images (or synthetic radiographs free of noise) using a ray-tracing algorithm implementing the python library TIGRE.

The second step is to `add' noise to the synthetic radiographs. The noise found in experimental radiographs does not follow a standard distribution such as a Gaussian function and therefore is difficult to simulate using traditional models. The noise can instead be applied to the synthetically produced noise-free radiographs using a conditional generative adversarial network (cGAN)~\cite{Cond}.
Similar to traditional GANs, a cGAN consists of a generator network and a discriminator network; however, rather than using a latent noise vector as the input of the generator, an image from experiment is used instead.
We use contrastive unpaired translation (CUT)~\cite{CUT} to model the noise found in the experimental radiographs. Figure~\ref{wrap-fig:cut} also shows the process of training CUT using synthetically produced radiographs.

The CUT model consists of a residual convolutional network~\cite{Residual} with nine residual blocks for the generator network $(G)$, PatchGAN~\cite{PatchGAN} for the discriminator network $(D)$ and a set of perceptrons $(H_l)$.
By using feature vectors from the $l$th layer of the generator's encoder $G_{enc}^l$, the perceptrons $(H_l)$ are used to classify image patches produced by network as patches from the original image.
Classification loss helps to inform the generator through optimization of the patch contrastive loss, by preserving semantic features of the synthetic data such as the location of shell boundaries.
The noise content of the experimental images is transferred to the synthetic images by jointly optimizing the generator and discriminator through the least squares GAN loss (LSGAN)~\cite{LSGAN}. 
The CUT model is trained using 200 synthetically produced images and 66 experimental images to generate the synthetic images shown in Figure ~\ref{fig:syn}.
%Figure ~\ref{fig:syn} compares images generated using CUT with experimental images.

\section{Super-resolution using neural network}
We recently demonstrated sub-pixel resolution or `super-resolution' using neural networks for post-processing of a boron-coated CCD (bCCD) pixelated images generated by neutrons. Similar results have also been obtained for data from CMOS sensors, which will be reported elsewhere.

The detection principle for ultracold neutrons (UCNs) using a bCCD was discussed previously. A scientific grade bCCD was used for UCN detection in our previous work \cite{kuk2021projection}. The bCCD sensor was built by the Lawrence Berkeley National Laboratory (LBNL) \cite{holland2003fully} and has been extensively characterized by Fermilab for the Dark Energy Camera (DECam) project \cite{flaugher2015dark}. The detector is a 250 $\mu$m thick, fully depleted, back-illuminated sensor fabriacated on high-resistivity silicon and has 8 million pixels (2k $\times$ 4k) with a pixel pitch of 15 $\times$ 15 $\mu$m$^2$. A thin $^{10}$B film up to 100 nm thick is deposited onto the transparent rear window of the bCCD camera to act as a conversion layer. The UCN hit is captured through the nuclear reactions $^{10}$B (n, $\alpha 0 \gamma$) $^7$Li (6\%) and $^{10}$B (n, $\alpha 1 \gamma$) $^7$Li (94\%). The charged particles $\alpha$, $^7$Li, and $\gamma$-rays, penetrates the active silicon layer of the detector to generate electron-hole (e-h) pairs. Influenced by the internal electric field, the generated holes will travel the full length of the active silicon layer to the potential wells near the poly electrode gates. The collected charges are converted to digitized value to be readout by the camera to create an output image of the UCN hit. 

\subsection{Allpix Squared}
High-statistics data samples produced with Monte Carlo simulations are required to train the neural network. The Allpix Squared semiconductor simulation framework~\cite{spannagel2018allpix, simon2} is used to generate these datasets. It is an open source simulation tool that implements end-to-end simulations of particle detection from incident radiation to digitized detector output. The framework comprises different algorithms for charge transport and front-end simulations as well as an interface to Geant4~\cite{agostinelli2003geant4} to describe the interaction of the incoming particle with the sensor material. Allpix Squared works on a first-principles basis, moving individual charge carriers or groups thereof along the electric field of the sensor using empirical mobility and recombination models. This approach allows to replicate the sensor response of imaging devices given the detector parameters and electric field distributions in the sensing element. Previous studies have demonstrated its capabilities of accurately describing the response of CMOS sensors to minimum ionizing particles~\cite{simon3}. 

The simulation is divided into several stages, each of which describes one component of the signal formation. In the first stage, the interaction of the incoming particle with the sensor material and the creation of electron-hole pairs is simulated. Subsequently, these charge carriers are propagated through the sensor in the second stage. The coupling to the front-end electronics is calculated in the third phase, and the front-end electronics and digitization is simulated in the fourth and last stage. For each of these stages, Allpix Squared can store the Monte Carlo truth information which allows to link detector output and initial particle and to trace the complete history of a detected pixel hit. This information can be exploited when training the neural network by providing both the true UCN position as well as the generated digital image obtained from the detector simulation. The built-in multithreading capabilities of Allpix Squared allow to scale the event generation and to simulate the large datasets required for training the neural network.

\subsection{bCCD Modeling in Allpix Squared}

While UCN hit images are acquired experimentally using a conversion layer and a silicon detector, the ground-truth UCN hit position is not available. However, we can obtain synthetically generated UCN hit images and their corresponding ground-truth hit position by using the silicon detector framework Allpix Squared as summarized above~\cite{spannagel2018allpix}. %Allpix Squared is an open source simulation framework for particle detection in silicon detectors. Built upon Geant4 \cite{agostinelli2003geant4}, Allpix Squared can simulate detector physics such as charge deposition and transport.
%In Allpix framework, the entire simulation history is recorded from beginning to end. Therefore, the ground-truth information such as the UCN hit position as well as the generated hit image can be easily obtained from the simulation history. 

\begin{figure}[!htb]
\centering
\includegraphics[width = 0.95\linewidth]{ 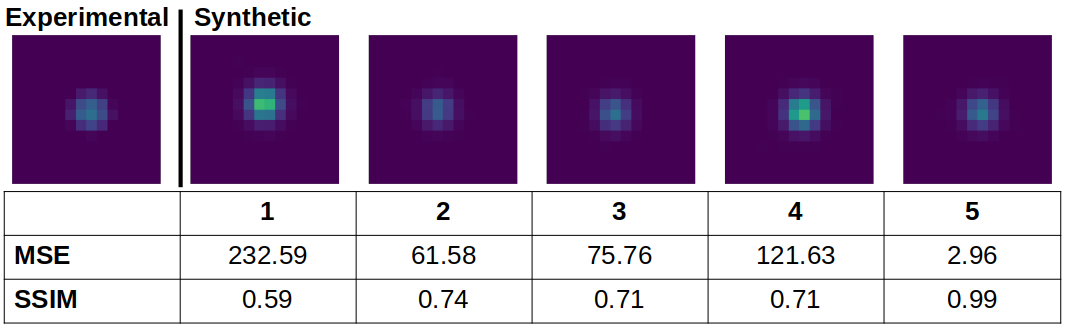}
\caption{We aim to match an experimental UCN hit with the best generated synthetic hit. The matching algorithm computes the mean squared error (MSE) between the experimental image and a synthetic image. The smaller the MSE, the closer the synthetic image is to the experimental. The structural similarity index measure (SSIM) between the experimental and synthetic image is also computed, where a perfect match corresponds to SSIM = 1. In this example, synthetic image 5 best matches the experimental.}
\label{fig:ucn_hit}
\end{figure}

To accurately model the silicon detector physics, Allpix requires the detector to be fully characterized so that the output physics of Allpix Squared can well match the actual detector. One physics check we perform is comparing the experimental UCN hit images with Allpix's synthetically generated images. Figure \ref{fig:ucn_hit} shows an example of matching an experimental hit with the closest generated synthetic image. We utilize a matching algorithm that computes the mean squared error (MSE) between the experimental and each synthetic image, and returns the synthetic image that results in the lowest MSE error. Recall that the generation of Allpix Squared hits is based on Monte Carlo simulations. Therefore, the more synthetic hits that are generated, the higher the chance of generating a synthetic hit that better matches the experimental.

\begin{figure}[tb]
\centering
\includegraphics[width = 0.95\linewidth]{ 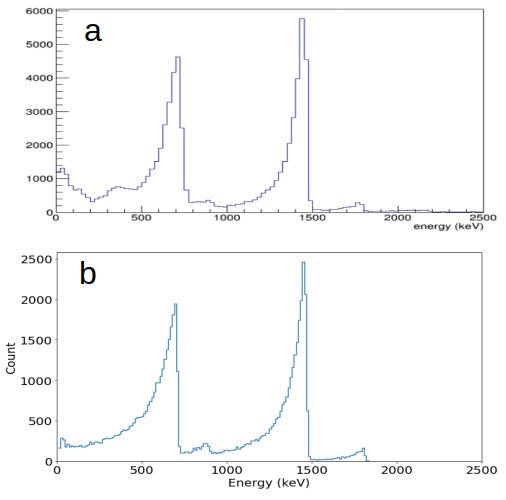}
\caption{The captured UCN energy spectrum by (a) the bCCD and (b) Allpix Squared. (Note: The expected peaks are $^7$Li = 840 keV and $\alpha$ =1470 keV.)}
\label{fig:spectrum}
\end{figure}

Another physics check is the verification of the captured UCN energy spectrum. Due to the nuclear reaction between a neutron and the $^{10}$B film, the detector will capture one of four possible particle and energy combinations as shown in Table \ref{tab:particle}. Figure \ref{fig:spectrum}a plots the captured energy spectrum of the experimental hits, while Figure \ref{fig:spectrum}b plots the Allpix spectrum. Both energy spectrum plots are reconstructed to properly center on the 1470 keV $\alpha$ peak. Under ideal conditions, the $^7$Li peak should naturally occur at 840 keV. However, the $^7$Li peak is shifted about 70 keV which motivates the existence of a dead layer between the $^{10}$B film and fully depleted silicon layer due to the down-shift in the energy spectrum peak. The Allpix spectrum was generated using the synthetic UCN hits while incorporating a dead layer into the bCCD model. The dead layer for the bCCD is fully characterized by LBNL \cite{groom2017quantum}. With the dead layer modeled, the Allpix energy spectrum distribution and peaks well matches the experimental, which shows that the energy loss and charge creation within the detector is well captured by Allpix.

\begin{table}[!htb]
\centering
\caption{Detection probability ($\omega^i$) and the produced reaction energy of the charged particles from the neutron capture process. Note that a dead layer exists between the $^{10}$B film and the fully depleted region of the Si sensor, which would reduce the actual energy captured.}
\label{tab:particle}
\begin{tabular}{l l l}
\hline
Ion & $\omega^i$ & Energy (keV) \\
\hline
$^7$Li & 47\% & 840 \\
$^7$Li & 3\% & 1020 \\
$\alpha$ & 47\% & 1470 \\
$\alpha$ & 3\% & 1780 \\
\hline
\end{tabular}

\end{table}

\subsection{Deep Learning for position super-resolution}

Machine learning techniques are very popular in recent years to learn a predictive model between input and output labels. Deep learning is a special case of machine learning that is very powerful in learning nonlinear predictive models by using neural networks \cite{lecun2015deep}. We aim to leverage deep learning to obtain a predictive model that maps from input UCN hit images to the ground-truth hit position. 

\begin{figure}[!htb]
\centering
\includegraphics[width=0.95\linewidth]{ 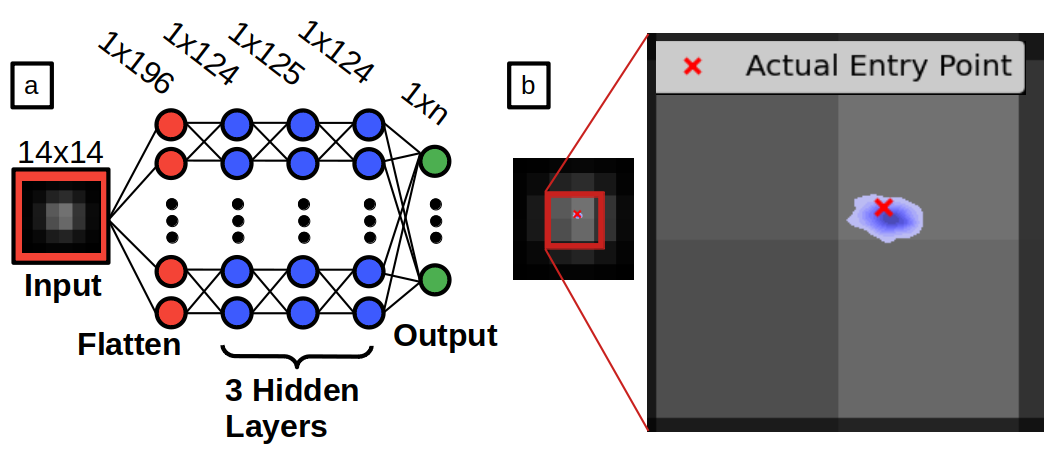}
\caption{(a) Overview of a FCNN model with three hidden layers. The input images of size $14\times14$ pixels are flattened into a 1-D vector. The output of the neural network is a 1-D vector of size $n$, which denotes the number of ground-truth labels. (b) The ground-truth labels in this example include the $(x,y)$ position for the hit entry point, where the red `x' denotes the actual entry point. The blue kernel density estimation (KDE) plot shows the FCNN prediction, which obtains sub-pixel position resolution.}
\label{fig:nn}
\end{figure}

Deep learning typically requires a large dataset to attain accurate predictive models. We use Allpix Squared to generate a large synthetic dataset consisting of 60,000 images and their corresponding ground-truth labels. Note that, in addition to the ground-truth hit position, other types of ground-truth information from the simulation history or prior simulation knowledge can be included in the output labels. The experimental UCN data is not used to train the neural network as the ground-truth labels are not available. Using the synthetic data, we propose to train a fully connected neural network (FCNN). Figure \ref{fig:nn}a shows the overview of an arbitrary FCNN model with three hidden layers. In the FCNN architecture, the 2-D input images are first flattened into a 1-D vector. The flattened layer is then followed by three hidden layers with 124, 125, and 124 hidden neurons, respectively, and the output layer. While not shown in Figure \ref{fig:nn}a, dropout layers are also included in the FCNN during the training and testing process. Dropout layers are popularly used in neural networks to help mitigate over-fitting issues as well as uncertainty quantification. We use the Pytorch library to train the FCNN to obtain a predictive mapping between the input UCN hit images and the ground-truth labels. The trained model can then be used to make predictions on the hit position of input UCN hit images, both synthetic and experimental images. Figure \ref{fig:nn}b shows an example FCNN prediction on the entry point position for a synthetic hit image with sub-pixel resolution. It is important to note that the accuracy of the predictive model for this arbitrary FCNN can be improved by tuning the network architecture, number of hidden layers and neurons, and other neural network parameters including the number of training epochs and learning rate.

\section{Summary}

A 20 Mfps CMOS image sensor design is described. TCAD simulations showed that the test chip can run at least 20 Mfps with less than 5.8 $e^-$ input-referred noise, the lowest noise reported in the ultra-fast burst-mode image sensor category. By using neural networks for post data processing, we demonstrated noise emulation and super position resolution at a fraction of pixel size. The combination of novel CMOS pixel designs and data-enabled image post-processing provide a promising path towards ultrafast multi-modal radiographic imaging and tomography applications.

SL and ZW wish to thank Drs. Don Groom and Steve Holland, both from Lawrence Berkeley National Laboratory, for stimulating discussions. This work is supported in part by the LANL LDRD, C3, and ICF programs under the Contract No. 89233218CNA000001. Prof. Zhu's group at UT Austin would like to acknowledge the NSF support through the Award 1802319.

\end{document}